\def\Ar{\rightarrow}
\def\bar{\overline}

\def\n{\nu}

\def\bar{\overline}
\def\l{\lambda}

\documentstyle [12 pt]{article}
\begin{document}
\setcounter{page}{1}
\thispagestyle{empty}
\topskip 0.5  cm
\vspace{1 cm}
\centerline{\LARGE \bf See-Saw Realization of the Texture Zeros}
\vskip 0.5 cm
\centerline{\LARGE \bf in the Neutrino Mass Matrix}
\vskip 1.5 cm
\centerline{{\large \bf Atsushi Kageyama}
 \renewcommand{\thefootnote}{\fnsymbol{footnote}}
\footnote[1]{E-mail address:  atsushi@muse.hep.sc.niigata-u.ac.jp},
\qquad{\large \bf Satoru Kaneko}
\renewcommand{\thefootnote}{\fnsymbol{footnote}}
\footnote[2]{E-mail address: kaneko@muse.hep.sc.niigata-u.ac.jp}}
\vskip 0.8 cm
\centerline{{\large \bf Noriyuki Shimoyama}
\renewcommand{\thefootnote}{\fnsymbol{footnote}}
\footnote[3]{E-mail address: simoyama@muse.hep.sc.niigata-u.ac.jp},
\qquad {\large \bf Morimitsu Tanimoto}
\renewcommand{\thefootnote}{\fnsymbol{footnote}}
\footnote[4]{E-mail address: tanimoto@muse.hep.sc.niigata-u.ac.jp}
 }
\vskip 0.8 cm
 \centerline{ \it{Department of Physics, Niigata University, 
 Ikarashi 2-8050, 950-2181 Niigata, JAPAN}}
\vskip 4 cm
\centerline{\bf ABSTRACT}\par
\vskip 1 cm

 We study the see-saw realization of seven textures of the neutrino
mass matrix, which were presented by Frampton, Glashow and Marfatia. 
Two of them ($\rm B_1$ and $\rm B_2$) are not realized in the see-saw
 mechanism without fine-tuning of parameters.
We present some specific textures of the Dirac neutrino mass matrix and 
the right-handed Majorana neutrino one. In order to test these textures,
we discuss the effect on the branching ratio of 
$\mu \rightarrow e \gamma$. 
  We also study the $U(1)_X \times U(1)_{X'}$  flavor symmetry, 
in which $U(1)_X$ is anomalous and  $U(1)_{X'}$ is  non-anomolous, 
 to  reproduce texture zeros.
 We present examples of U(1) charges for two textures  
  ($\rm A_1$ and $\rm A_2$) .
 
\newpage
\topskip 0. cm

 The Super-Kamiokande has almost confirmed the neutrino oscillation
 in the atmospheric neutrinos, which favors the $\n_\mu\Ar \nu_\tau$
process  \cite{SKam}.
For the solar neutrinos \cite{SKamsolar,SNO}, the recent data of 
the Super-Kamiokande and the Sudbury Neutrino Observatory 
 also favor strongly the neutrino oscillation
$\n_e\Ar \nu_{\mu,\tau}$ with the large mixing angle(LMA) MSW solution 
\cite{MSW,Lisi}. 
These results indicate the neutrino masses and mixings,    
especially, the bi-large flavor mixing. 
It is therefore important to investigate how the textures of lepton mass 
matrices can link up with the observables of the flavor mixing \cite{MNS,FT}.

 Recently,  Frampton, Glashow and Marfatia \cite{Fram} found seven
acceptable textures of the neutrino mass matrix with two independent
vanishing entries in the basis with the diagonal charged lepton masses. 
 The further study of these textures  
was presented by Xing \cite{Xing}.  Another insight has been also given
focusing on texture zeros \cite{Kang}.

Since these textures are given for the light effective neutrino mass matrix,
 one needs to find the see-saw realization \cite{Seesaw}
of these textures from the standpoint of the model building.
In this paper we have examined the see-saw realization of those seven
textures.  It is found that two of them are not given by the see-saw mechanism
without fine tunings between parameters  of the Dirac neutrino
mass matrix  and the right-handed Majorana neutrino one.
Some specific textures of the Dirac neutrino mass matrix and the right-handed
Majorana neutrino one are presented.
These textures are discussed in terms of the branching ratio of 
$\mu \rightarrow e \gamma$ in order to test them experimentally. 
 The $U(1)_X \times U(1)_{X'}$ flavor symmetry, in which 
 $U(1)_X$ is anomalous and  $U(1)_{X'}$ is  non-anomolous, is discussed to 
reproduce texture zeros. In this symmetry, zeros in the mass matrix are 
derived from SYSY zeros, which are due to holomorphy of the superpotential.

There are seven acceptable textures with two independent zeros
for the effective neutrino mass matrix as shown in Table 1.
The cases $\rm A_1$ and $\rm A_2$ correspond to the hierarchical
neutrino mass spectrum while  $\rm B_1 \sim B_4$ and $\rm C$  correspond to 
the degenerate neutrino mass spectrum.
Generally these textures are realized in the see-saw mechanism.
However, as far as we exclude the possibility that these zeros are originated
 from accidental cancellations in the see-saw mechanism, 
the see-saw realization of these seven textures are not trivial.
Then, these zeros should come from  zeros of the Dirac neutrino mass matrix 
and the right-handed Majorana mass matrix.
Therefore it is significant to study the see-saw realization
of these seven textures. 

In terms of the Dirac neutrino mass matrix ${\bf m_D}$ and
the right-handed Majorana neutrino mass matrix ${\bf M_R}$,
the effective neutrino mass matrix ${\bf M^\nu}$ is given as 
\begin{equation}
\bf M^\nu=  m_D\   M_R^{\rm -1}\  m_D^{\rm T} \ .
\end{equation} 
\noindent Therefore,  
each entry of the effective neutrino mass matrix ${\bf M^\nu}$
is given in terms of  $m_{ij}$, which is the component of ${\bf m_D}$, and
$M_{ij}$, which is the component of ${\bf M_R}$.
Zeros in $m_{ij}$  and $M_{ij}$ provide texture zeros
in the effective neutrino mass matrix since we do not consider accidental 
cancellations between $m_{ij}$  and $M_{ij}$.
The components of the effective neutrino mass matrix are written as
\begin{equation}
  M_{ij}^{\nu}= \sum_{k,\ \ell=1}^3\ m_{ik}\ m_{j\ell}\ 
({\bf M_R^{\rm -1}})_{k\ell}\ ,
\label{see}
\end{equation} 
\noindent where
\begin{equation}
{\bf M_R^{\rm -1}}=\frac{1}{D}
 \left (\matrix{M_{22}M_{33}-M_{23}^2 & M_{13}M_{23}-M_{12}M_{33} &
  M_{12}M_{23}-M_{13}M_{22} \cr
M_{13}M_{23}-M_{12}M_{33} & M_{11}M_{33}-M_{13}^2 & M_{12}M_{13}-M_{11}M_{23}
 \cr
M_{12}M_{23}-M_{13}M_{22} & M_{12}M_{13}-M_{11}M_{23} & M_{11}M_{22}-M_{12}^2
 \cr  } \right)\ , 
\end{equation} 
\noindent where  $D$ denotes the determinant of ${\bf M_R}$.
In order to answer the question  where zeros come from,
we try to find   textures with  zeros for the Dirac neutrino mass matrix
 and the right-handed Majorana neutrino mass matrix by
solving $M_{ij}^{\nu}=0$ without cancellations among terms in eq.(\ref{see}).
This has been done by both analytic search and computer search.

 At first, we present three remarks, in which the texture zeros are not
obtained by the see-saw mechanism.

 Remark 1 :  There is no see-saw realization in the cases of 
$\rm B_1$ and $\rm B_2$. 

 For the case $\rm B_1$,  putting constraints 
$M^\nu_{13}=M^\nu_{31}=M^\nu_{22}=0$ with other non-zero  components,
we have searched textures with  zeros of the Dirac and 
the right-handed Majorana neutrino mass matrices
without fine tuning of the parameters.  
It is concluded that there is no solution to reproduce the case $\rm B_1$.
There is also no solution for $\rm B_2$.
Thus,  $\rm B_1$ and $\rm B_2$ are unfavored 
 for the model building with the see-saw mechanism.
Therefore, the following studies focus on the five textures 
$\rm A_1,\ A_2, \ B_3, \ B_4$ and $\rm C$.

 Remark 2 : There is no see-saw realization for the five textures
 unless both Dirac neutrino mass matrix and right-handed Majorana neutrino 
one have zeros.  In other words, the texture of $\bf m_D$ or $\bf M_R$ 
without zeros does not lead to desired five textures
$\rm A_1,\ A_2, \ B_3, \ B_4$ and $\rm C$.

Remark 3 : There is no see-saw realization for the five textures
 if the right-handed Majorana neutrino mass matrix is diagonal.
 This remark is important one 
since there are many discussions taking the diagonal
basis for the right-handed Majorana neutrino mass matrix.

These  remarks are useful guides to build models of the
neutrino mass matrix.
 
If the Dirac neutrino mass matrix  is specified,
 we can find possible textures of the right-handed Majorana neutrinos.
In order to see the situation, we  consider the Fritzsch texture 
in the Dirac neutrino mass matrix \cite{Fri} for simplicity.
In this case, there are three textures of the right-handed Majorana neutrinos
 only for the type $\rm A_2$ as follows:
\begin{equation}
{\bf m_D}=\left ( \matrix{
{\bf 0} & \times &  {\bf 0}\cr
 \times &  {\bf 0} &  \times\cr
 {\bf 0} & \times & \times \cr} \right ) \ ; 
\quad
{\bf M_R}=\left ( \matrix{
{\bf 0} & \times &  {\bf 0}\cr
 \times &  {\bf 0} &  \times \cr
 {\bf 0} & \times  & \times \cr} \right ), 
\quad
\left ( \matrix{
{\bf 0} & \times &  {\bf 0}\cr
 \times &  \times &   {\bf 0}\cr
 {\bf 0} &  {\bf 0}  & \times \cr} \right )  ,
\quad
\left ( \matrix{
{\bf 0} & \times &  {\bf 0}\cr
 \times &   {\bf 0}& {\bf 0}\cr
 {\bf 0} &  {\bf 0}  & \times \cr} \right )  ,
\end{equation}
\noindent
where $\times$ denotes non-zero entry.
For other four textures $\rm A_1, \ B_3, \ B_4$ and $\rm C$, 
 there is no solution. Thus, the type of 
 ${\bf M^\nu}$  and the texture of the right-handed Majorana neutrinos
 is selected once the texture of the Dirac neutrinos is specified.

Another simple example is the diagonal texture for the Dirac neutrino mass
matrix. Then, the right-handed Majorana neutrino mass matrices are
shown in Table 1.  There are only two textures for $\rm A_1$, $\rm A_2$,
$\rm B_3$ and $\rm B_4$, respectively,  but nothing for $\rm C$.

 Now, let us fix the right-handed Majorana neutrino mass matrix.  
The first attempt is the simplest texture, 
  which is parametrized by two parameters.
This case corresponds to at least 
two degenerate right-handed Majorana neutrino masses.
There are three types ${\rm a_0}$,  ${\rm b_0}$,  ${\rm c_0}$ as follows:

\begin{equation}
{\bf M_R}\ ; \quad
 {\rm a_0}: \quad \left ( \matrix{
{\bf 0} &   {\bf 0}& \times\cr
   {\bf 0} & \times & {\bf 0}\cr
 \times & {\bf 0} & {\bf 0} \cr} \right ) \ ,
\quad
 {\rm b_0}: \quad\left ( \matrix{
{\bf 0} & \times &  {\bf 0}\cr
 \times &  {\bf 0} &  {\bf 0}\cr
 {\bf 0} & {\bf 0} & \times \cr} \right )\ , 
\quad
 {\rm c_0}: \quad \left ( \matrix{
\times & {\bf 0} &  {\bf 0}\cr
  {\bf 0} &  {\bf 0} &  \times\cr
 {\bf 0} & \times &  {\bf 0} \cr} \right )\ . 
\qquad
\label{Majo2}
\end{equation}
\noindent
 It is helpful to comment on the symmetry of these texture.
 The textures ${\rm b_0}$ and ${\rm c_0}$ are derived
 through the permutation of flavors of the right-handed neutrinos
 in  the texture ${\rm a_0}$.
 The permutation of $2 \leftrightarrow 3$  and the cyclic permutation  
 $3\rightarrow 2 \rightarrow 1 \rightarrow 3$ in ${\rm a_0}$
give  ${\rm b_0}$,  while the 
$1\leftrightarrow 2$ and the cyclic permutation 
$1\rightarrow 2 \rightarrow 3 \rightarrow 1$  in ${\rm a_0}$
give  ${\rm c_0}$.

For ${\rm a_0}$, ${\rm b_0}$ and ${\rm c_0}$ of  ${\bf M_R}$,
 there are ten textures of the Dirac neutrino mass matrix
in $\rm A_1,\ A_2, \ B_3$ and  $\rm B_4$, respectively 
while only two textures in $\rm C$.
In Tables 2 - 6, we present typical textures of the Dirac neutrino mass 
matrix for ${\rm a_0}$ and ${\rm b_0}$.
As seen in the tables, textures of the Dirac neutrino mass 
matrix for ${\rm b_0}$ are obtained 
by the permutation  $2\leftrightarrow 3$ in the right-handed sector
(columuns in the Dirac neutrino mass matrix). 
The cyclic permutation  
 $3\rightarrow 2 \rightarrow 1 \rightarrow 3$ also leads to
another texture of the Dirac neutrino mass matrix, which is not
presented in the tables. 

Let us discuss about the ten textures of the Dirac neutrinos for cases
  $\rm A_1,\ A_2, \ B_3$ and  $\rm B_4$.
Two of them have three independent zeros, six of them have four zeros 
and remained two textures have five zeros.
 For $\rm C$, there are only two textures with four zeros.
 We show these textures taking  ${\rm a_0}$  for the right-handed 
Majorana neutrino in the case of  $\rm A_1$:

\begin{equation}
{\bf m_D}= \left ( \matrix{ {\bf 0} & {\bf 0}& \times\cr
   {\bf 0} & \times & \times \cr  \times & \times & \times \cr} \right )  ,
\ 
 \left ( \matrix{ {\bf 0} & {\bf 0}& \times\cr
   {\bf 0} & \times & \times \cr  \times &  {\bf 0} & \times \cr} \right )  ,
\ 
 \left ( \matrix{ {\bf 0} & {\bf 0}& \times\cr
   {\bf 0} & \times &  {\bf 0} \cr  \times & \times & \times \cr} \right )  ,
\ 
 \left ( \matrix{ {\bf 0} & {\bf 0}& \times\cr
   {\bf 0} & \times & \times \cr  \times & \times &  {\bf 0}\cr} \right )  ,
\
 \left ( \matrix{ {\bf 0} & {\bf 0}& \times\cr
 {\bf 0} & \times &  {\bf 0} \cr  \times & \times &  {\bf 0} \cr} \right )  ,
\label{example}
\end{equation}
\noindent
 and other five textures are obtained by the symmetry such as:
\begin{equation}
{\bf m'_D}={\bf m_D \ P}  \ , \quad  {\rm with} \quad
 {\bf P} = \left (\matrix{ 0 & 0& 1\cr 0 & 1 & 0\cr 1& 0 & 0 \cr} \right ) \ .
\end{equation}

Since the case ${\rm c_0}$  is  also obtained from  ${\rm a_0}$ 
by the permutation  $1\leftrightarrow 2$ and
the cyclic permutation $1\rightarrow 2 \rightarrow 3 \rightarrow 1$ 
in the right-handed sector, we have omitted results of ${\rm c_0}$ in the 
tables. 

The right-handed Majorana neutrino mass matrices in eq.(\ref{Majo2})
 are modified by adding a new parameter. 
Since these have three independent zeros,
the mass eigenvalues are not degenerate in general.
The texture $\rm a_0$ is modified as follow:

\begin{equation}
 {\bf M_R}\ ; \  {\rm a_1}: \ \left ( \matrix{
\times &   {\bf 0}& \times\cr
   {\bf 0} & \times & {\bf 0}\cr
 \times & {\bf 0} & {\bf 0} \cr} \right ) ,
\quad {\rm a_2}: \ 
 \left ( \matrix{
{\bf 0} &  \times& \times\cr
  \times & \times & {\bf 0}\cr
 \times & {\bf 0} & {\bf 0} \cr} \right ) ,
\quad {\rm a_3}: \ 
 \left ( \matrix{
{\bf 0} &   {\bf 0}& \times\cr
   {\bf 0} & \times & {\bf 0}\cr
 \times & {\bf 0} & \times \cr} \right ) ,
\quad {\rm a_4}: \ 
 \left ( \matrix{
{\bf 0} &   {\bf 0}& \times\cr
   {\bf 0} & \times & \times\cr
 \times & \times & {\bf 0} \cr} \right )  .
\end{equation}
\noindent
For the texture  $\rm b_0$, we have 

\begin{equation}
{\bf M_R}\ ;  \  {\rm b_1}:\ \left ( \matrix{
 \times & \times &  {\bf 0}\cr
 \times &  {\bf 0} &  {\bf 0}\cr
 {\bf 0} & {\bf 0} & \times \cr} \right ), 
\ \  {\rm b_2}: \ 
\left ( \matrix{
{\bf 0} & \times &  \times \cr
 \times &  {\bf 0} &  {\bf 0}\cr
 \times  & {\bf 0} & \times \cr} \right ), 
\ \  {\rm b_3}: \ 
\left ( \matrix{
{\bf 0} & \times &  {\bf 0}\cr
 \times &  \times &  {\bf 0}\cr
 {\bf 0} & {\bf 0} & \times \cr} \right ), 
\ \  {\rm b_4}: \
\left ( \matrix{
{\bf 0} & \times &  {\bf 0}\cr
 \times &  {\bf 0} &  \times\cr
 {\bf 0} & \times & \times \cr} \right )  ,
\end{equation}
\noindent and 
for the texture  $\rm c_0$, we have 

\begin{equation}
{\bf M_R}\ ; \ {\rm c_1}:\
 \left ( \matrix{
\times &   {\bf 0}&  {\bf 0}\cr
   {\bf 0} & \times & \times\cr
 {\bf 0} & \times & {\bf 0} \cr} \right ) ,
\quad  {\rm c_2}:\ \left ( \matrix{
 \times & \times &  {\bf 0}\cr
 \times &  {\bf 0} &  \times\cr
 {\bf 0} & \times  & {\bf 0}\cr} \right ), 
\quad {\rm c_3}:\
 \left ( \matrix{
\times &   {\bf 0}&  {\bf 0}\cr
   {\bf 0} &  {\bf 0} & \times\cr
 {\bf 0} & \times & \times \cr} \right ) ,
\quad {\rm c_4}: \left ( \matrix{
\times & {\bf 0} &  \times\cr
  {\bf 0} &  {\bf 0} &  \times\cr
 \times & \times &  {\bf 0} \cr} \right ) \ .
\end{equation}

For textures of ${\bf M_R}$ with one zero in the diagonal entries
(${\rm a_1},{\rm a_3},{\rm b_1},{\rm b_3}$),
there are 6 textures of the Dirac neutrino mass matrix to reproduce 
$\rm A_1, \ A_2, \  B_3$ and $\rm  B_4$.  
For textures of ${\bf M_R}$ with two zeros
in the diagonal entries (${\rm a_2},{\rm a_4},{\rm b_2},{\rm b_4}$),
there are 8 textures of the Dirac neutrino mass matrix 
for $\rm A_1,\ A_2, \ B_3$ and $\rm B_4$.
  It is remarked that there is no texture of the Dirac neutrino mass matrix to
reproduce $\rm C$.
We show the typical textures of the Dirac neutrino mass matrix
in Tables 2 - 6.
 Since the ${\rm b_i}\ ({\rm i}=1-4)$ are also derived by the permutation  
$2\leftrightarrow 3$  in  ${\rm a_i}$, the textures of the Dirac neutrino 
mass matrix correspond to 
 the permutation  $2\leftrightarrow 3$ in their columuns. 
 The cyclic permutation  $3\rightarrow 2 \rightarrow 1 \rightarrow 3$
also  leads to another texture of the Dirac neutrino mass matrix. 
 We have also omitted the case of ${\rm c_i}$  in tables. 

What kind of experiments can test these textures? 
It is well known that the Yukawa coupling of the neutrino contributes to
 the lepton flavor violation (LFV).  
Many authors have studied the LFV in the minimal supersymmetric standard 
model (MSSM) with right-handed neutrinos assuming the relevant neutrino 
mass matrix \cite{Borz,LFV1,LFV2,Sato,Casas,our}.
In the MSSM with soft breaking terms, 
there exist  lepton flavor violating terms such as 
off-diagonal elements of slepton mass matrices 
and trilinear couplings (A-term).
It is noticed that large neutrino Yukawa couplings and large lepton mixings 
 generate the large LFV in the left-handed slepton masses. For example, 
the decay rate of  $\mu\rightarrow e \gamma$ can be approximated as follows:

\begin{eqnarray}
{\rm \Gamma} (\mu\rightarrow e \gamma) \simeq 
\frac{e^2}{16 \pi} m^5_{\mu} F
 \left| \frac{(6+2 a_0^2)m_0^2}{16 \pi^2}
({\bf Y_\nu  Y_\nu^\dagger})_{21} \ln\frac{M_X}{M_R}  \right|^2\ ,
\label{rate}
\end{eqnarray}
\noindent where the neutrino Yukawa coupling matrix  $\bf Y_\nu$ 
 is given as  ${\bf Y_\nu}={\bf m_D}/v_2$ ($v_2$ is a VEV of Higgs)
at the right-handed mass scale $M_R$, and 
  $F$ is a function of masses and mixings for SUSY particles.  
In eq.(\ref{rate}),
 we assume the universal scalar mass $(m_0)$ for all scalars and
the universal A-term $(A_f=a_0 m_0 Y_f)$ at the GUT scale $M_X$. 
Therefore  the branching ratio $\mu \rightarrow e \gamma$ depends considerably on the texture of the Dirac neutrino \cite{Sato,Casas,our}.

Let us investigate  ${\bf m_D m_D^\dagger}$ focusing 
on the process $\mu\rightarrow e \gamma$.
Zeros in the Dirac neutrino mass matrix may suppress the
 $\mu\rightarrow e \gamma$ decay.
There are  many textures, which lead to $({\bf m_D m_D^\dagger})_{21}=0$
as seen in eq.(\ref{example}).
 Since the contribution of the neutrino Yukawa couplings on 
 $\mu\rightarrow e \gamma$  is tiny in these textures,
 the branching ratio is safely predicted to be below the present experimental
 upper bound $1.2\times 10^{-11}$ \cite{exp}. We call these textures
as type II and other textures as type I, which can contribute
considerably to the branching ratio.
We present  typical  textures of the Dirac neutrino mass matrix  and
those numbers by classifying the type I and type II in Tables 2, 3, 4 and 5.
The type I and II will be tested in the future experiments of the
  $\mu\rightarrow e \gamma$ decay.

Thus, we find a lot of sets which lead to five textures 
$\rm A_1, \ A_2$, $\rm B_3, \ B_4$ and $\rm C$ 
through the see-saw mechanism without fine-tuning.
The important problem is the origin of zeros.
An interesting origin of zeros is the holomorphic zeros in the supersymmetry 
with the anomalous U(1) flavor symmetry \cite{U1}, 
which are called SUSY zeros.
Although one can impose SUSY zeros in the neutrino mass matrix by
assigning the relavant U(1) charges, one cannot choose the
diagonal basis in the charged lepton mass matrix.
In other words, it is impossible to impose zeros 
of these textures keeping the diagonal charged lepton mass matrix in the case of the U(1) flavor symmetry.

A simple way to avoid this difficulty is to have one anomalous $U(1)_X$
and another non-anomolous  $U(1)_{X'}$, which are broken by two scalar fields $\phi_1$ and $\phi_2$ having the flavor charges $(-1,-1)$ and $(0,1)$ 
\cite{Choi}. 
Since the D-term potential guarantees 
$\langle\phi_1\rangle=\langle\phi_2\rangle$ in this model,
the effective interactions of the relevant sector are given as
\begin{equation}         
    L_i \bar \ell_j H_1 \l^{x_i+y_j-x'_i-y'_j}\  \l^{x_i+y_j} +
    \frac{1}{M_R} L_i  L_j H_2 H_2 \l^{x_i+x_j-x'_i-x'_j} \ 
   \l^{x_i+x_j}\ ,
\label{Int}
\end{equation}
  \noindent where
\begin{equation}         
  \l\equiv \frac{\langle\phi_1\rangle}{\Lambda}=
 \frac{\langle\phi_2\rangle}{\Lambda} \ ,
\end{equation}
\noindent
 and $(x_i,x'_i)$ are flavor charges of left-handed lepton doublets
$L_i$, and $(y_j,y'_j)$ are flavor charges of right-handed lepton singlets
$\ell_j$ for $U(1)_X$ and  $U(1)_{X'}$, respectively.
If the power of $\l$ in eq.(\ref{Int}) is negative, the interaction
is forbidden. 
Then, the texture zero, which is SUSY zero, is reproduced 
in the mass matrix.

We find  U(1) charges to realize the patterns of $\rm A_1$ and $\rm A_2$
as follows:
\begin{eqnarray}         
 {\rm A_1} : &&  (x_1, x_2, x_3)=(-3,2,4)\ , \quad
            (x'_1, x'_2, x'_3)=(-11,0,4)\ ,  \nonumber \\
             &&  (y_1, y_2, y_3)=(3,0,-4)\ , \quad
            (y'_1, y'_2, y'_3)=(6,2,-4)\ ,
\label{NA1}
\end{eqnarray}
\noindent which  give
\begin{equation}         
  \bf M^\nu \propto {\rm \l^8} \left ( \matrix{ 0  &  0& \l \cr
    0 & 1 & 1\cr  \l  & 1 & 1 \cr} \right ) \ , \qquad
 \bf M^e \propto \left ( \matrix{ \l^5  &  0 & 0 \cr
    0 & \l^2 & 0 \cr  0 & 0  & 1 \cr} \right ) \ , 
\end{equation}
\noindent and
\begin{eqnarray}         
 {\rm A_2} : &&  (x_1, x_2, x_3)=(-3,4,2)\ , \quad
            (x'_1, x'_2, x'_3)=(-11,4,0)\ ,  \nonumber \\
             &&  (y_1, y_2, y_3)=(3,-4,-2)\ , \quad
            (y'_1, y'_2, y'_3)=(6,-6,0)\ ,
\label{NA2}
\end{eqnarray}
\noindent which give
\begin{equation}         
  \bf M^\nu \propto {\rm \l^8} \left ( \matrix{ 0  &  \l & 0 \cr
    \l  & 1 & 1\cr  0  & 1 & 1 \cr} \right ) \ , \qquad
 \bf M^e \propto \left ( \matrix{ \l^5  &  0 & 0 \cr
    0 & \l^2 & 0 \cr  0 & 0  & 1 \cr} \right ) \ , 
\end{equation}
 where  U(1) charges of Higgs scalars are  taken to be zero,
and $\l\simeq 0.22$ is supposed.
Since the magnitude of each entry has been estimated numerically by Xing 
\cite{Xing}, we can compare eqs.(\ref{NA1}) and (\ref{NA2}) 
with his numerical results.     
We have found that our results are completely  consistent with  them.
Since the  degenerate mass spectrum is unlikely realized
in the U(1) flavor symmetry, we do not discuss the U(1) charges
 in the pattern of  $\rm B_3, \ B_4$ and $\rm C$. 


 In order to present the see-saw realization of the U(1) symmetry,
we show two examples including the U(1) charges of right-handed
neutrinos.
 For  ${\rm A_1}$,   U(1) charges of right-handed
  neutrinos $N_i$ and $N'_i$ are taken  as
\begin{equation}         
             (N_1, N_2, N_3)=(3,-3,0)    \ , \qquad
             (N'_1, N'_2, N'_3)=(6,-6,-5) \ ,  
\end{equation}
\noindent which  with eq.(\ref{NA1}) give
\begin{equation}         
  \bf M_R \propto  \left ( \matrix{ 0  &  1 & \l^5 \cr
    1 & 0 & 0\cr  \l^5  & 0 & \l^{10} \cr} \right ) \ , \qquad
 \bf m_D  \propto {\rm \l^4} \left ( \matrix{ \l  &  0 & 0 \cr
    0 & 0  & \l^5 \cr  0 & 1  & \l^5 \cr} \right ) \ .
\end{equation}
\noindent
For  ${\rm A_2}$,  we take
\begin{equation}         
             (N_1, N_2, N_3)=(3,-3,0)    \ , \qquad
             (N'_1, N'_2, N'_3)=(4, -4, -1) \ ,  
\end{equation}
\noindent which  with eq.(\ref{NA2}) give
\begin{equation}         
  \bf M_R \propto \left ( \matrix{ 0  &  1 & \l^3 \cr
    1 & 0 & 0\cr  \l^3  & 0 & \l^2 \cr} \right ) \ , \qquad
 \bf m_D \propto  {\rm \l^2} \left ( \matrix{ \l^5  &  0 & 0 \cr
    0 & 1  & \l^3 \cr  \l^4 & 0  & \l^3 \cr} \right ) \ .
\end{equation}

Thus, the $U(1)_X \times U(1)_{X'}$ flavor symmetry gives the neutrino mass 
matrix which is consistent with the experimental data.

 The summary is given as follows.
 We have studied the see-saw realization of seven textures of the neutrino
mass matrix, which were presented by Frampton, Glashow and Marfatia. 
Two of them ($\rm B_1$ and $\rm B_2$) are not realized in the see-saw
 mechanism without fine-tuning of parameters.
Some specific textures of the Dirac neutrino mass matrix and the right-handed
Majorana neutrino one have been presented.
In order to test these textures experimentally, 
 the branching ratio of $\mu \rightarrow e \gamma$ has been discussed. 
  
The $U(1)_X \times U(1)_{X'}$ flavor symmetry, in which one $U(1)_X$
is anomalous  and  $U(1)_{X'}$ is  non-anomolous, has been  studied to 
reproduce the texture zeros.
 An example of U(1) charges has been found for $\rm A_1$ and $\rm A_2$,
respectively, which give hierarchical neutrino masses.
Other approaches  are also necessary  
in the model building of two-zeros textures \cite{FramX}.

 It is expected that the texture of the neutrino mass matrix
 is more constrained by the further comming experiments.
 Therefore, the systematic study becomes also available in the framework of 
the see-saw mechanism in the near future.
\vskip 1  cm
 
 We would like to thank Dr. S.K. Kang for useful discussions.
 This research is supported by the Grant-in-Aid for Science Research,
 Ministry of Education, Science and Culture, Japan(No.12047220). 

\newpage

\newpage
\begin{table}
\caption{Seven patterns of the neutrino mass matrix ${\bf M^\nu}$ with 
two independent vanishing entries. 
Possible right-handed Majorana mass matrices
are presented in the case of the diagonal Dirac neutrino mass matrix.}
\begin{center}
\begin{tabular}{ccccccccc} \hline \hline \\
Pattern &~& Texture of $\bf M^\nu$ &~& Dirac ${\bf m_D}$ &~~~& 
  Majorana $\bf M_R$ &~& \\ \\ \hline \\

$\rm A_1$
&&
 $\left ( \matrix{
{\bf 0} & {\bf 0} & \times \cr
{\bf 0} & \times & \times \cr
\times & \times & \times \cr} \right )$
&&
 $\left ( \matrix{
\times & {\bf 0} & {\bf 0} \cr
{\bf 0} & \times & {\bf 0} \cr
{\bf 0} & {\bf 0} & \times \cr} \right )$
&~~~&
 $\left ( \matrix{
\times & \times & \times \cr
\times & \times & {\bf 0} \cr
\times & {\bf 0} & {\bf 0} \cr} \right )$
\hspace{-0.2cm}$\matrix{\cr\cr,\cr}$\hspace{-0.2cm}
 $\left ( \matrix{
{\bf 0} & \times & \times \cr
\times & \times & {\bf 0} \cr
\times & {\bf 0} & {\bf 0} \cr} \right )$
\\ \\ \hline\\
$\rm A_2$
&& 
$\left ( \matrix{
{\bf 0} & \times & {\bf 0} \cr
\times & \times & \times \cr
{\bf 0} & \times & \times \cr} \right )$
&&
 $\left ( \matrix{
\times & {\bf 0} & {\bf 0} \cr
{\bf 0} & \times & {\bf 0} \cr
{\bf 0} & {\bf 0} & \times \cr} \right )$
&~~~&
 $\left ( \matrix{
\times & \times & \times \cr
\times & {\bf 0} & {\bf 0} \cr
\times & {\bf 0} & \times \cr} \right )$
\hspace{-0.2cm}$\matrix{\cr\cr,\cr}$\hspace{-0.2cm}
 $\left ( \matrix{
{\bf 0} & \times & \times \cr
\times & {\bf 0} & {\bf 0} \cr
\times & {\bf 0} & \times \cr} \right )$

\\ \\  \hline\\
$\rm B_1$
&& 
$\left ( \matrix{
\times & \times & {\bf 0} \cr
\times & {\bf 0} & \times \cr
{\bf 0} & \times & \times \cr} \right )$
&&
---------------------------
&&
------------------------------------
\\ \\ \hline\\
$\rm B_2$
&& 
$\left ( \matrix{
\times & {\bf 0} & \times \cr
{\bf 0} & \times & \times \cr
\times & \times & {\bf 0} \cr} \right )$
&&
---------------------------
&&
------------------------------------
\\ \\ \hline\\
$\rm B_3$
&&
 $\left ( \matrix{
\times & {\bf 0} & \times \cr
{\bf 0} & {\bf 0} & \times \cr
\times & \times & \times \cr} \right )$
&&
 $\left ( \matrix{
\times & {\bf 0} & {\bf 0} \cr
{\bf 0} & \times & {\bf 0} \cr
{\bf 0} & {\bf 0} & \times \cr} \right )$
&~~~&
 $\left ( \matrix{
\times & \times & {\bf 0} \cr
\times & \times & \times \cr
{\bf 0} & \times & {\bf 0} \cr} \right )$
\hspace{-0.2cm}$\matrix{\cr\cr,\cr}$\hspace{-0.2cm}
 $\left ( \matrix{
\times & \times & {\bf 0} \cr
\times & {\bf 0} & \times \cr
{\bf 0} & \times & {\bf 0} \cr} \right )$
\\ \\ \hline \\
$\rm B_4$
&&
 $\left ( \matrix{
\times & \times & {\bf 0} \cr
\times & \times & \times \cr
{\bf 0} & \times & {\bf 0} \cr} \right )$
&&
 $\left ( \matrix{
\times & {\bf 0} & {\bf 0} \cr
{\bf 0} & \times & {\bf 0} \cr
{\bf 0} & {\bf 0} & \times \cr} \right )$
&~~~&
 $\left ( \matrix{
\times & {\bf 0} & \times \cr
{\bf 0} & {\bf 0} & \times \cr
\times & \times & \times \cr} \right )$
\hspace{-0.2cm}$\matrix{\cr\cr,\cr}$\hspace{-0.2cm}
 $\left ( \matrix{
\times & {\bf 0} & \times \cr
{\bf 0} & {\bf 0} & \times \cr
\times & \times & {\bf 0} \cr} \right )$
\\ \\ \hline \\
$\rm C$
&&
 $\left ( \matrix{
\times & \times & \times \cr
\times & {\bf 0} & \times \cr
\times & \times & {\bf 0} \cr} \right )$
&&
 $\left ( \matrix{
\times & {\bf 0} & {\bf 0} \cr
{\bf 0} & \times & {\bf 0} \cr
{\bf 0} & {\bf 0} & \times \cr} \right )$
&&
nothing
\\ \\ \hline\hline
\end{tabular}
\end{center}
\end{table}
\normalsize


\begin{table}
\caption{ In the case of ${\rm  A_1}$, 
typical Dirac neutrino textures of types I and II  are presented  
for the right-handed Majorana neutrino textures
 $\rm a_i$ and $\rm b_i$.
Textures $\rm c_i$ are omitted. The total  $\sharp $ denotes  
the numbers of possible  Dirac neutrino textures,
 $\sharp $ denotes the numbers of  Dirac neutrino textures of types I and II,
respectively.}
\begin{center}
\begin{tabular}{cccccccccccccccc} \hline \hline 
& & Majorana ${\bf M_R}$ && total $\sharp$ && Typical $\bf m_D$ &&
 $\sharp$ && Typical $\bf m_D $ && $\sharp $ &&
\\&& && && of type I &&  && of typeII && &&\\  
\hline \vspace{- 0.3 cm}\\
& ${\rm a}_0$ &
 $\left ( \matrix{
{\bf 0} & {\bf 0} & \times \cr
{\bf 0} & \times & {\bf 0} \cr
\times & {\bf 0} & {\bf 0} \cr} \right )$
&&
10
&&
 $\left ( \matrix{
{\bf 0} & {\bf 0} & \times \cr
{\bf 0} & \times & \times \cr
\times & \times & \times \cr} \right )$
&&
6
&&
 $\left ( \matrix{
{\bf 0} & {\bf 0} & \times \cr
{\bf 0} & \times & {\bf 0} \cr
\times & \times & \times \cr} \right )$
&&
4
\vspace{0.2 cm}\\ \hline\vspace{-0.3 cm}\\
& ${\rm a}_1$ &
 $\left ( \matrix{
\times & {\bf 0} & \times \cr
{\bf 0}& \times & {\bf 0} \cr
\times & {\bf 0} & {\bf 0} \cr} \right )$
&&
6
&&
 $\left ( \matrix{
\times & {\bf 0} & {\bf 0} \cr
\times & \times & {\bf 0} \cr
\times & \times & \times \cr} \right )$
&&
4
&&
 $\left ( \matrix{
\times  & {\bf 0} & {\bf 0} \cr
{\bf 0} & \times & {\bf 0} \cr
\times & \times & \times \cr} \right )$
&&
2
\vspace{0.2 cm}\\ \hline\vspace{-0.3 cm}\\
& ${\rm a}_2$ & 
$\left ( \matrix{
{\bf 0} & \times & \times \cr
\times & \times & {\bf 0} \cr
\times & {\bf 0}  & {\bf 0} \cr} \right )$
&&
8
&&
 $\left ( \matrix{
\times & {\bf 0} & {\bf 0} \cr
\times & \times & {\bf 0} \cr
\times & \times & \times \cr} \right )$
&&
4
&&
 $\left ( \matrix{
\times  & {\bf 0} & {\bf 0} \cr
{\bf 0} & \times & {\bf 0} \cr
\times & \times & \times \cr} \right )$
&&
4
\vspace{0.2 cm}\\ \hline\vspace{-0.3 cm}\\
& ${\rm a}_3$ & 
 $\left ( \matrix{
{\bf 0}& {\bf 0} & \times \cr
{\bf 0} & \times & {\bf 0} \cr
\times & {\bf 0} & \times \cr} \right )$
&&
6
&&
 $\left ( \matrix{
{\bf 0} & {\bf 0} & \times \cr
{\bf 0} & \times & \times \cr
\times & \times & \times \cr} \right )$
&&
4
&&
 $\left ( \matrix{
{\bf 0}  & {\bf 0} & \times \cr
{\bf 0} & \times & {\bf 0} \cr
\times & \times & \times \cr} \right )$
&&
2
\vspace{0.2 cm}\\ \hline\vspace{-0.3 cm}\\
& ${\rm a}_4$ &
 $\left ( \matrix{
{\bf 0}& {\bf 0} & \times \cr
{\bf 0}& \times & \times \cr
\times & \times & {\bf 0} \cr} \right )$
&&
8
&&
 $\left ( \matrix{
{\bf 0} & {\bf 0} & \times \cr
{\bf 0} & \times & \times \cr
\times & \times & \times \cr} \right )$
&&
4
&&
 $\left ( \matrix{
{\bf 0}  & {\bf 0} & \times \cr
{\bf 0} & \times & {\bf 0} \cr
\times & \times & \times \cr} \right )$
&&
4
\vspace{0.2 cm}\\ \hline\vspace{-0.3 cm}\\

& ${\rm b}_0$ & 
 $\left ( \matrix{
{\bf 0} & \times & {\bf 0} \cr
\times & {\bf 0} & {\bf 0} \cr
{\bf 0} & {\bf 0} & \times \cr} \right )$
&&
10
&&
 $\left ( \matrix{
{\bf 0} & \times & {\bf 0} \cr
{\bf 0} & \times & \times \cr
\times & \times & \times \cr} \right )$
&&
6
&&
 $\left ( \matrix{
{\bf 0} & \times & {\bf 0} \cr
{\bf 0} & {\bf 0} & \times \cr
\times & \times & \times \cr} \right )$
&&
4
\vspace{0.2 cm}\\ \hline\vspace{-0.3 cm}\\
& ${\rm b}_1$ &
 $\left ( \matrix{
\times & \times & {\bf 0} \cr
\times & {\bf 0} & {\bf 0} \cr
{\bf 0} & {\bf 0} & \times \cr} \right )$
&&
6
&&
 $\left ( \matrix{
\times & {\bf 0} & {\bf 0} \cr
\times & {\bf 0} & \times \cr
\times & \times & \times \cr} \right )$
&&
4
&&
 $\left ( \matrix{
\times  & {\bf 0} & {\bf 0} \cr
{\bf 0} & {\bf 0} & \times \cr
\times & \times & \times \cr} \right )$
&&
2
\vspace{0.2 cm}\\ \hline\vspace{-0.3 cm}\\
& ${\rm b}_2$ &
 $\left ( \matrix{
{\bf 0} & \times & \times \cr
\times & {\bf 0} & {\bf 0} \cr
\times & {\bf 0} & \times \cr} \right )$
&&
8
&&
 $\left ( \matrix{
\times & {\bf 0} & {\bf 0} \cr
\times & {\bf 0} & \times \cr
\times & \times & \times \cr} \right )$
&&
4
&&
 $\left ( \matrix{
\times  & {\bf 0} & {\bf 0} \cr
{\bf 0} & {\bf 0} & \times \cr
\times & \times & \times \cr} \right )$
&&
4
\vspace{0.2 cm}\\ \hline\vspace{-0.3 cm}\\
& ${\rm b}_3$ &
 $\left ( \matrix{
{\bf 0} & \times & {\bf 0} \cr
\times & \times & {\bf 0} \cr
{\bf 0} & {\bf 0} & \times \cr} \right )$
&&
6
&&
 $\left ( \matrix{
{\bf 0} & \times & {\bf 0} \cr
{\bf 0} & \times & \times \cr
\times & \times & \times \cr} \right )$
&&
4
&&
 $\left ( \matrix{
{\bf 0}  & \times & {\bf 0} \cr
{\bf 0} & {\bf 0} & \times \cr
\times & \times & \times \cr} \right )$
&&
2
\vspace{0.2 cm}\\ \hline\vspace{-0.3 cm}\\
& ${\rm b}_4$ &
 $\left ( \matrix{
{\bf 0}& \times & {\bf 0} \cr
\times & {\bf 0} & \times \cr
{\bf 0} & \times & \times \cr} \right )$
&&
8
&&
 $\left ( \matrix{
{\bf 0} & \times & {\bf 0} \cr
{\bf 0} & \times & \times \cr
\times & \times & \times \cr} \right )$
&&
4
&&
 $\left ( \matrix{
{\bf 0}  & \times & {\bf 0} \cr
{\bf 0} & {\bf 0} & \times \cr
\times & \times & \times \cr} \right )$
&&
4
\vspace{0.2 cm}\\ \hline\hline
\end{tabular}
\end{center}
\end{table}

\newpage
\begin{table}
\caption{ In the case of ${\rm A_2}$,
typical Dirac neutrino textures of types I and II  are presented.}
\begin{center}
\begin{tabular}{cccccccccccccccc} \hline \hline 
& & Majorana ${\bf M_R}$ && total $\sharp$ && Typical $\bf m_D$ &&
 $\sharp$ && Typical $\bf m_D $ && $\sharp $ &&
\\&& && && of type I &&  && of typeII && &&\\  
\hline \vspace{- 0.3 cm}\\
& ${\rm a}_0$ &
 $\left ( \matrix{
{\bf 0} & {\bf 0} & \times \cr
{\bf 0} & \times & {\bf 0} \cr
\times & {\bf 0} & {\bf 0} \cr} \right )$
&&
10
&&
 $\left ( \matrix{
{\bf 0} & {\bf 0} & \times \cr
\times & \times & \times \cr
{\bf 0} & \times & \times \cr} \right )$
&&
6
&&
 $\left ( \matrix{
{\bf 0} & {\bf 0} & \times \cr
\times & \times & {\bf 0} \cr
{\bf 0} & \times & \times \cr} \right )$
&&
4
\vspace{0.2 cm}\\ \hline\vspace{-0.3 cm}\\
& ${\rm a}_1$ &
 $\left ( \matrix{
\times & {\bf 0} & \times \cr
{\bf 0}& \times & {\bf 0} \cr
\times & {\bf 0} & {\bf 0} \cr} \right )$
&&
6
&&
 $\left ( \matrix{
\times & {\bf 0} & {\bf 0} \cr
\times & \times & \times \cr
\times & \times & {\bf 0} \cr} \right )$
&&
3
&&
 $\left ( \matrix{
\times & {\bf 0} & {\bf 0} \cr
{\bf 0} & \times & \times \cr
\times & \times & {\bf 0} \cr} \right )$
&&
3
\vspace{0.2 cm}\\ \hline\vspace{-0.3 cm}\\
& ${\rm a}_2$ & 
$\left ( \matrix{
{\bf 0} & \times & \times \cr
\times & \times & {\bf 0} \cr
\times & {\bf 0}  & {\bf 0} \cr} \right )$
&&
8
&&
 $\left ( \matrix{
\times & {\bf 0} & {\bf 0} \cr
\times & \times & \times \cr
\times & \times & {\bf 0} \cr} \right )$
&&
4
&&
 $\left ( \matrix{
\times & {\bf 0} & {\bf 0} \cr
{\bf 0} & \times & \times \cr
\times & \times & {\bf 0} \cr} \right )$
&&
4
\vspace{0.2 cm}\\ \hline\vspace{-0.3 cm}\\
& ${\rm a}_3$ & 
 $\left ( \matrix{
{\bf 0}& {\bf 0} & \times \cr
{\bf 0} & \times & {\bf 0} \cr
\times & {\bf 0} & \times \cr} \right )$
&&
6
&&
 $\left ( \matrix{
{\bf 0} & {\bf 0} & \times \cr
\times & \times & \times \cr
{\bf 0} & \times & \times \cr} \right )$
&&
3
&&
 $\left ( \matrix{
{\bf 0} & {\bf 0} & \times \cr
\times & \times & {\bf 0} \cr
{\bf 0} & \times & \times \cr} \right )$
&&
3
\vspace{0.2 cm}\\ \hline\vspace{-0.3 cm}\\
& ${\rm a}_4$ &
 $\left ( \matrix{
{\bf 0}& {\bf 0} & \times \cr
{\bf 0}& \times & \times \cr
\times & \times & {\bf 0} \cr} \right )$
&&
8
&&
 $\left ( \matrix{
{\bf 0} & {\bf 0} & \times \cr
\times & \times & \times \cr
{\bf 0} & \times & \times \cr} \right )$
&&
4
&&
 $\left ( \matrix{
{\bf 0} & {\bf 0} & \times \cr
\times & \times & {\bf 0} \cr
{\bf 0} & \times & \times \cr} \right )$
&&
4
\vspace{0.2 cm}\\ \hline\vspace{-0.3 cm}\\

& ${\rm b}_0$ & 
 $\left ( \matrix{
{\bf 0} & \times & {\bf 0} \cr
\times & {\bf 0} & {\bf 0} \cr
{\bf 0} & {\bf 0} & \times \cr} \right )$
&&
10
&&
 $\left ( \matrix{
{\bf 0} & \times & {\bf 0} \cr
\times & \times & \times \cr
{\bf 0} & \times & \times \cr} \right )$
&&
6
&&
 $\left ( \matrix{
{\bf 0} & \times & {\bf 0} \cr
\times & {\bf 0} & \times \cr
{\bf 0} & \times & \times \cr} \right )$
&&
4
\vspace{0.2 cm}\\ \hline\vspace{-0.3 cm}\\
& ${\rm b}_1$ &
 $\left ( \matrix{
\times & \times & {\bf 0} \cr
\times & {\bf 0} & {\bf 0} \cr
{\bf 0} & {\bf 0} & \times \cr} \right )$
&&
6
&&
 $\left ( \matrix{
\times & {\bf 0} & {\bf 0} \cr
\times & \times & \times \cr
\times & {\bf 0} & \times \cr} \right )$
&&
3
&&
 $\left ( \matrix{
\times & {\bf 0} & {\bf 0} \cr
{\bf 0} & \times & \times \cr
\times & {\bf 0} & \times \cr} \right )$
&&
3
\vspace{0.2 cm}\\ \hline\vspace{-0.3 cm}\\
& ${\rm b}_2$ &
 $\left ( \matrix{
{\bf 0} & \times & \times \cr
\times & {\bf 0} & {\bf 0} \cr
\times & {\bf 0} & \times \cr} \right )$
&&
8
&&
 $\left ( \matrix{
\times & {\bf 0} & {\bf 0} \cr
\times & \times & \times \cr
\times & {\bf 0} & \times \cr} \right )$
&&
4
&&
 $\left ( \matrix{
\times & {\bf 0} & {\bf 0} \cr
{\bf 0} & \times & \times \cr
\times & {\bf 0} & \times \cr} \right )$
&&
4
\vspace{0.2 cm}\\ \hline\vspace{-0.3 cm}\\
& ${\rm b}_3$ &
 $\left ( \matrix{
{\bf 0} & \times & {\bf 0} \cr
\times & \times & {\bf 0} \cr
{\bf 0} & {\bf 0} & \times \cr} \right )$
&&
6
&&
 $\left ( \matrix{
{\bf 0} & \times & {\bf 0} \cr
\times & \times & \times \cr
{\bf 0} & \times & \times \cr} \right )$
&&
3
&&
 $\left ( \matrix{
{\bf 0} & \times & {\bf 0} \cr
\times & {\bf 0} & \times \cr
{\bf 0} & \times & \times \cr} \right )$
&&
3
\vspace{0.2 cm}\\ \hline\vspace{-0.3 cm}\\
& ${\rm b}_4$ &
 $\left ( \matrix{
{\bf 0}& \times & {\bf 0} \cr
\times & {\bf 0} & \times \cr
{\bf 0} & \times & \times \cr} \right )$
&&
8
&&
 $\left ( \matrix{
{\bf 0} & \times & {\bf 0} \cr
\times & \times & \times \cr
{\bf 0} & \times & \times \cr} \right )$
&&
4
&&
 $\left ( \matrix{
{\bf 0} & \times & {\bf 0} \cr
\times & {\bf 0} & \times \cr
{\bf 0} & \times & \times \cr} \right )$
&&
4
\vspace{0.2 cm}\\ \hline\hline
\end{tabular}
\end{center}
\end{table}

\newpage

\begin{table}
\caption{ In the case of ${\rm  B_3}$, 
typical Dirac neutrino textures of types I and II  are presented.}
\begin{center}
\begin{tabular}{cccccccccccccccc} \hline \hline 
& & Majorana ${\bf M_R}$ && total $\sharp$ && Typical $\bf m_D$ &&
 $\sharp$ && Typical $\bf m_D $ && $\sharp $ &&
\\&& && && of type I &&  && of typeII && &&\\  
\hline \vspace{- 0.3 cm}\\

& ${\rm a}_0$ &
 $\left ( \matrix{
{\bf 0} & {\bf 0} & \times \cr
{\bf 0} & \times & {\bf 0} \cr
\times & {\bf 0} & {\bf 0} \cr} \right )$
&&
10
&&
 $\left ( \matrix{
{\bf 0} & \times & \times \cr
{\bf 0} & {\bf 0} & \times \cr
\times & \times & \times \cr} \right )$
&&
6
&&
 $\left ( \matrix{
{\bf 0} & \times & {\bf 0} \cr
{\bf 0} & {\bf 0} & \times \cr
\times & \times & \times \cr} \right )$
&&
4
\vspace{0.2 cm}\\ \hline\vspace{-0.3 cm}\\
& ${\rm a}_1$ &
 $\left ( \matrix{
\times & {\bf 0} & \times \cr
{\bf 0}& \times & {\bf 0} \cr
\times & {\bf 0} & {\bf 0} \cr} \right )$
&&
6
&&
 $\left ( \matrix{
\times & \times & {\bf 0} \cr
\times & {\bf 0} & {\bf 0} \cr
\times & \times & \times \cr} \right )$
&&
4
&&
 $\left ( \matrix{
{\bf 0} & \times & {\bf 0} \cr
\times & {\bf 0} & {\bf 0} \cr
\times & \times & \times \cr} \right )$
&&
2
\vspace{0.2 cm}\\ \hline\vspace{-0.3 cm}\\
& ${\rm a}_2$ & 
$\left ( \matrix{
{\bf 0} & \times & \times \cr
\times & \times & {\bf 0} \cr
\times & {\bf 0}  & {\bf 0} \cr} \right )$
&&
8
&&
 $\left ( \matrix{
\times & \times& {\bf 0} \cr
\times & {\bf 0} & {\bf 0} \cr
\times & \times & \times \cr} \right )$
&&
4
&&
 $\left ( \matrix{
{\bf 0} & \times & {\bf 0} \cr
\times & {\bf 0} & {\bf 0} \cr
\times & \times & \times \cr} \right )$
&&
4
\vspace{0.2 cm}\\ \hline\vspace{-0.3 cm}\\
& ${\rm a}_3$ & 
 $\left ( \matrix{
{\bf 0}& {\bf 0} & \times \cr
{\bf 0} & \times & {\bf 0} \cr
\times & {\bf 0} & \times \cr} \right )$
&&
6
&&
 $\left ( \matrix{
{\bf 0} & \times & \times \cr
{\bf 0} & {\bf 0} & \times \cr
\times & \times & \times \cr} \right )$
&&
4
&&
 $\left ( \matrix{
{\bf 0} & \times & {\bf 0} \cr
{\bf 0} & {\bf 0} & \times \cr
\times & \times & \times \cr} \right )$
&&
2
\vspace{0.2 cm}\\ \hline\vspace{-0.3 cm}\\
& ${\rm a}_4$ &
 $\left ( \matrix{
{\bf 0}& {\bf 0} & \times \cr
{\bf 0}& \times & \times \cr
\times & \times & {\bf 0} \cr} \right )$
&&
8
&&
 $\left ( \matrix{
{\bf 0} & \times & \times \cr
{\bf 0} & {\bf 0} & \times \cr
\times & \times & \times \cr} \right )$
&&
4
&&
 $\left ( \matrix{
{\bf 0} & \times & {\bf 0} \cr
{\bf 0} & {\bf 0} & \times \cr
\times & \times & \times \cr} \right )$
&&
4
\vspace{0.2 cm}\\ \hline\vspace{-0.3 cm}\\

& ${\rm b}_0$ & 
 $\left ( \matrix{
{\bf 0} & \times & {\bf 0} \cr
\times & {\bf 0} & {\bf 0} \cr
{\bf 0} & {\bf 0} & \times \cr} \right )$
&&
10
&&
 $\left ( \matrix{
{\bf 0} & \times & \times \cr
{\bf 0} & \times & {\bf 0} \cr
\times & \times & \times \cr} \right )$
&&
6
&&
 $\left ( \matrix{
{\bf 0} & {\bf 0} & \times \cr
{\bf 0} & \times & {\bf 0} \cr
\times & \times & \times \cr} \right )$
&&
4
\vspace{0.2 cm}\\ \hline\vspace{-0.3 cm}\\
& ${\rm b}_1$ &
 $\left ( \matrix{
\times & \times & {\bf 0} \cr
\times & {\bf 0} & {\bf 0} \cr
{\bf 0} & {\bf 0} & \times \cr} \right )$
&&
6
&&
 $\left ( \matrix{
\times & {\bf 0} & \times \cr
\times & {\bf 0} & {\bf 0} \cr
\times & \times & \times \cr} \right )$
&&
4
&&
 $\left ( \matrix{
{\bf 0} & {\bf 0} & \times \cr
\times & {\bf 0} & {\bf 0} \cr
\times & \times & \times \cr} \right )$
&&
2
\vspace{0.2 cm}\\ \hline\vspace{-0.3 cm}\\
& ${\rm b}_2$ &
 $\left ( \matrix{
{\bf 0} & \times & \times \cr
\times & {\bf 0} & {\bf 0} \cr
\times & {\bf 0} & \times \cr} \right )$
&&
8
&&
 $\left ( \matrix{
\times & {\bf 0} & \times \cr
\times & {\bf 0} & {\bf 0} \cr
\times & \times & \times \cr} \right )$
&&
4
&&
 $\left ( \matrix{
{\bf 0} & {\bf 0} & \times \cr
\times & {\bf 0} & {\bf 0} \cr
\times & \times & \times \cr} \right )$
&&
4
\vspace{0.2 cm}\\ \hline\vspace{-0.3 cm}\\
& ${\rm b}_3$ &
 $\left ( \matrix{
{\bf 0} & \times & {\bf 0} \cr
\times & \times & {\bf 0} \cr
{\bf 0} & {\bf 0} & \times \cr} \right )$
&&
6
&&
 $\left ( \matrix{
{\bf 0} & \times & \times \cr
{\bf 0} & \times & {\bf 0} \cr
\times & \times & \times \cr} \right )$
&&
4
&&
 $\left ( \matrix{
{\bf 0} & {\bf 0} & \times \cr
{\bf 0} & \times & {\bf 0} \cr
\times & \times & \times \cr} \right )$
&&
2
\vspace{0.2 cm}\\ \hline\vspace{-0.3 cm}\\
& ${\rm b}_4$ &
 $\left ( \matrix{
{\bf 0}& \times & {\bf 0} \cr
\times & {\bf 0} & \times \cr
{\bf 0} & \times & \times \cr} \right )$
&&
8
&&
 $\left ( \matrix{
{\bf 0} & \times & \times \cr
{\bf 0} & \times & {\bf 0} \cr
\times & \times & \times \cr} \right )$
&&
4
&&
 $\left ( \matrix{
{\bf 0} & {\bf 0} & \times \cr
{\bf 0} & \times & {\bf 0} \cr
\times & \times & \times \cr} \right )$
&&
4
\vspace{0.2 cm}\\ \hline\hline
\end{tabular}
\end{center}
\end{table}

\begin{table}
\caption{ In the case of ${\rm  B_4}$, 
typical Dirac neutrino textures of types I and II  are presented.}
\begin{center}
\begin{tabular}{cccccccccccccccc} \hline \hline 
& & Majorana ${\bf M_R}$ && total $\sharp$ && Typical $\bf m_D$ &&
 $\sharp$ && Typical $\bf m_D $ && $\sharp $ &&
\\&& && && of type I &&  && of typeII && &&\\  
\hline \vspace{- 0.3 cm}\\
& ${\rm a}_0$ &
 $\left ( \matrix{
{\bf 0} & {\bf 0} & \times \cr
{\bf 0} & \times & {\bf 0} \cr
\times & {\bf 0} & {\bf 0} \cr} \right )$
&&
10
&&
 $\left ( \matrix{
\times & \times & {\bf 0} \cr
\times & \times & \times \cr
\times & {\bf 0} & {\bf 0} \cr} \right )$
&&
10
&&
 nothing
\vspace{0.2 cm}\\ \hline\vspace{-0.3 cm}\\
& ${\rm a}_1$ &
 $\left ( \matrix{
\times & {\bf 0} & \times \cr
{\bf 0}& \times & {\bf 0} \cr
\times & {\bf 0} & {\bf 0} \cr} \right )$
&&
6
&&
 $\left ( \matrix{
\times & \times & {\bf 0} \cr
\times & \times & \times \cr
\times & {\bf 0} & {\bf 0} \cr} \right )$
&&
5
&&
 $\left ( \matrix{
\times & \times & {\bf 0} \cr
{\bf 0} & {\bf 0} & \times \cr
\times & {\bf 0} & {\bf 0} \cr} \right )$
&&
1
\vspace{0.2 cm}\\ \hline\vspace{-0.3 cm}\\
& ${\rm a}_2$ & 
$\left ( \matrix{
{\bf 0} & \times & \times \cr
\times & \times & {\bf 0} \cr
\times & {\bf 0}  & {\bf 0} \cr} \right )$
&&
8
&&
 $\left ( \matrix{
\times & \times & {\bf 0} \cr
\times & \times & \times \cr
\times & {\bf 0} & {\bf 0} \cr} \right )$
&&
5
&&
 $\left ( \matrix{
\times & \times & {\bf 0} \cr
{\bf 0} & {\bf 0} & \times \cr
\times & {\bf 0} & {\bf 0} \cr} \right )$
&&3
\vspace{0.2 cm}\\ \hline\vspace{-0.3 cm}\\
& ${\rm a}_3$ & 
 $\left ( \matrix{
{\bf 0}& {\bf 0} & \times \cr
\times & \times & {\bf 0} \cr
{\bf 0} & {\bf 0} & \times \cr} \right )$
&&
6
&&
 $\left ( \matrix{
{\bf 0} & \times & \times \cr
\times & \times & \times \cr
{\bf 0} & {\bf 0} & \times \cr} \right )$
&&
5
&&
 $\left ( \matrix{
{\bf 0} & \times & \times \cr
\times & {\bf 0} & {\bf 0} \cr
{\bf 0} & {\bf 0} & \times \cr} \right )$
&&
1
\vspace{0.2 cm}\\ \hline\vspace{-0.3 cm}\\
& ${\rm a}_4$ &
 $\left ( \matrix{
{\bf 0}& {\bf 0} & \times \cr
{\bf 0}& \times & \times \cr
\times & \times & {\bf 0} \cr} \right )$
&&
8
&&
 $\left ( \matrix{
{\bf 0} & \times & \times \cr
\times & \times & \times \cr
{\bf 0} & {\bf 0} & \times \cr} \right )$
&&
5
&&
 $\left ( \matrix{
{\bf 0} & \times & \times \cr
\times & {\bf 0} & {\bf 0} \cr
{\bf 0} & {\bf 0} & \times \cr} \right )$
&&
3
\vspace{0.2 cm}\\ \hline\vspace{-0.3 cm}\\

& ${\rm b}_0$ & 
 $\left ( \matrix{
{\bf 0} & \times & {\bf 0} \cr
\times & {\bf 0} & {\bf 0} \cr
{\bf 0} & {\bf 0} & \times \cr} \right )$
&&
10
&&
 $\left ( \matrix{
\times & {\bf 0} & \times  \cr
\times & \times & \times \cr
\times & {\bf 0} & {\bf 0} \cr} \right )$
&&
10
&&
 nothing
\vspace{0.2 cm}\\ \hline\vspace{-0.3 cm}\\
& ${\rm b}_1$ &
 $\left ( \matrix{
\times & \times & {\bf 0} \cr
\times & {\bf 0} & {\bf 0} \cr
{\bf 0} & {\bf 0} & \times \cr} \right )$
&&
6
&&
 $\left ( \matrix{
\times & {\bf 0} & \times \cr
\times & \times & \times \cr
\times & {\bf 0} & {\bf 0} \cr} \right )$
&&
5
&&
 $\left ( \matrix{
\times & {\bf 0} & \times \cr
{\bf 0} & \times & {\bf 0} \cr
\times & {\bf 0} & {\bf 0} \cr} \right )$
&&
1
\vspace{0.2 cm}\\ \hline\vspace{-0.3 cm}\\
& ${\rm b}_2$ &
 $\left ( \matrix{
{\bf 0} & \times & \times \cr
\times & {\bf 0} & {\bf 0} \cr
\times & {\bf 0} & \times \cr} \right )$
&&
8
&&
 $\left ( \matrix{
\times & {\bf 0} & \times \cr
\times & \times & \times \cr
\times & {\bf 0} & {\bf 0} \cr} \right )$
&&
5
&&
 $\left ( \matrix{
\times & {\bf 0} & \times \cr
{\bf 0} & \times & {\bf 0} \cr
\times & {\bf 0} & {\bf 0} \cr} \right )$
&&
3
\vspace{0.2 cm}\\ \hline\vspace{-0.3 cm}\\
& ${\rm b}_3$ &
 $\left ( \matrix{
{\bf 0} & \times & {\bf 0} \cr
\times & \times & {\bf 0} \cr
{\bf 0} & {\bf 0} & \times \cr} \right )$
&&
6
&&
 $\left ( \matrix{
{\bf 0} & \times & \times \cr
\times & \times & \times \cr
{\bf 0} & \times & {\bf 0} \cr} \right )$
&&
5
&&
 $\left ( \matrix{
{\bf 0} & \times & \times \cr
\times & {\bf 0} & {\bf 0} \cr
{\bf 0} & \times & {\bf 0} \cr} \right )$
&&
1
\vspace{0.2 cm}\\ \hline\vspace{-0.3 cm}\\
& ${\rm b}_4$ &
 $\left ( \matrix{
{\bf 0}& \times & {\bf 0} \cr
\times & {\bf 0} & \times \cr
{\bf 0} & \times & \times \cr} \right )$
&&
8
&&
 $\left ( \matrix{
{\bf 0} & \times & \times \cr
\times & \times & \times \cr
{\bf 0} & \times & {\bf 0} \cr} \right )$
&&
5
&&
 $\left ( \matrix{
{\bf 0} & \times & \times \cr
\times & {\bf 0} & {\bf 0} \cr
{\bf 0} & \times & {\bf 0} \cr} \right )$
&&
3
\vspace{0.2 cm}\\ \hline\hline
\end{tabular}
\end{center}
\end{table}


\begin{table}
\caption{ In the case of ${\rm C}$,  typical Dirac neutrino textures 
of type I  are presented.  
In this case,  there is no type II and  
there are 2  textures of the Dirac neutrino mass matrix
 only  for  $\rm a_0$ and $\rm b_0$.}
\begin{center}
\begin{tabular}{cccccccccccccccc} \hline \hline 
& & Majorana ${\bf M_R}$ && total $\sharp$ && Typical $\bf m_D$ &&
 $\sharp$ && Typical $\bf m_D $ && $\sharp $ &&
\\&& && && of type I &&  && of type II && &&\\  
\hline \vspace{- 0.3 cm}\\
& ${\rm a}_0$ &
 $\left ( \matrix{
{\bf 0} & {\bf 0} & \times \cr
{\bf 0} & \times & {\bf 0} \cr
\times & {\bf 0} & {\bf 0} \cr} \right )$
&&
2
&&
 $\left ( \matrix{
\times & \times & \times \cr
\times & {\bf 0} & {\bf 0} \cr
{\bf 0} & {\bf 0} & \times \cr} \right )$
&&
2
&&
 nothing
\vspace{0.2 cm}\\ \hline\vspace{-0.3 cm}\\

& ${\rm b}_0$ & 
 $\left ( \matrix{
{\bf 0} & \times & {\bf 0} \cr
\times & {\bf 0} & {\bf 0} \cr
{\bf 0} & {\bf 0} & \times \cr} \right )$
&&
2
&&
 $\left ( \matrix{
  \times & \times & \times \cr
  \times & {\bf 0} & {\bf 0} \cr
 {\bf 0} & \times & {\bf 0} \cr} \right )$
&&
2
&&
 nothing

\vspace{0.2 cm}\\ \hline\hline
\end{tabular}
\end{center}
\end{table}
\end{document}